\documentclass[fleqn,twoside,twocolumn,11pt]{article}
\usepackage{graphicx}
\newcommand{\AmS}{{\protect\the\textfont2
  A\kern-.1667em\lower.5ex\hbox{M}\kern-.125emS}}

\title{%
Full one-loop electroweak radiative corrections to %
single photon production in $e^+e^-$  }

\newsavebox{\abstractbox}
\author{F. Boudjema${}^{a,1}$, %
J. Fujimoto${}^{b}$, %
T. Ishikawa${}^{b}$, T. Kaneko${}^{b}$, %
K. Kato${}^{c,2}$, %
Y.Kurihara${}^{b}$, \\ Y. Shimizu${}^{b}$, %
S. Yamashita${}^{d}$, %
Y. Yasui${}^{b}$ \\
${}^{a}$ {\it \small LAPTH, B.P.110, Annecy-le-Vieux F-74941, France} \\
${}^{b}$ {\it \small KEK, Oho 1-1, Tsukuba, Ibaraki 305-0801, Japan} \\
${}^{c}$ {\it \small Kogakuin University, Nishi-Shinjuku 1-24, %
Shinjuku, Tokyo 163-8677, Japan} \\
${}^{d}$ {\it \small ICEPP, University of Tokyo, %
Hongo 7-3-1, Bunkyo, Tokyo 113-0033, Japan} \\ 
{ } \\
{\bf Abstract} \\
\usebox{\abstractbox}
}
\date{}
      
\begin{document}
%%\maketitle

\sbox{\abstractbox}{\begin{minipage}{0.8\textwidth}
%%\begin{abstract}
Large scale calculation for the radiative corrections required
for the current and future collider experiments can be done
automatically using the GRACE-LOOP system.
Here several results for $e^+e^-\rightarrow$ 3-body processes
are presented including $ e^+e^-\rightarrow e^+e^-H$ and
$ e^+e^-\rightarrow \nu\bar{\nu}\gamma$.
\vspace{1pc}
%%\end{abstract}
\end{minipage}}

\maketitle


\addtocounter{footnote}{1}
\footnotetext{URA 14-36 du CNRS, %
Associ\'{e}e \`{a} l'Universit\'{e} de Savoie.}
\addtocounter{footnote}{1}
\footnotetext{Presented by K.Kato in ACAT03.}

\section{Introduction}

The research of the Higgs particle is one of the
most important subjects in the particle physics.
Its discovery is never be the final target but the
detailed study of its property is the key to understand
the standard model and its possible extension.
The same refers to expected 
supersymmetric(SUSY) particles.
Experimental study will be done in the future $e^+e^-$ linear colliders(LC).
In order to improve the theoretical prediction
for the experimental data with high accuracy,
the electroweak(EW) radiative corrections are required for
various signal channels and major background ones.

Armored by the automated systems, now the calculation is
available not only for the two-to-two processes
but for the two-to-three processes. Already,
the analyses of the following 
three processes are found in the literature.

The important Higgs production processes in LC are the 
Higgsstrahlung $e^+e^-\rightarrow ZH$ and
the $W$-fusion process $e^+e^-\rightarrow \nu_e\bar{\nu}_eH$.
The latter is dominant in the large parameter space.
The tree level calculation shows that the $W$-fusion
is dominant at $W=800\mathrm{GeV}$ for $M_H\le 600\mathrm{GeV}$
and also even at $W=500\mathrm{GeV}$ for Higgs in 
the mass-range preferred by precision data study\cite{EW}.
The full EW correction for this process is done in
\cite{nunuH} and has demonstrated the importance of the complete
calculation. The resulted weak correction is $-2\sim -4\mathrm{\%}$
in $G_{\mu}$-scheme.

The direct study of Higgs Yukawa coupling can be done in
LC for the production channel $e^+e^-\rightarrow t\bar{t}H$
whose cross section is order of $\mathrm{fb}$ for a light Higgs
for $W=700\sim 1000\mathrm{GeV}$. The QCD correction has been
studied\cite{ttHQCD} and the correction is negative in the high
energy region. The full EW correction is studied in \cite{ttH}
and has shown that the correction is similar magnitude as
that of QCD and is positive in the high energy region.

More challenging item is the study of Higgs self coupling
(Higgs potential). Though the yield is low, $0.2\mathrm{fb}$
for $W\simeq 500\mathrm{GeV}$ and $M_H\simeq 120\mathrm{GeV}$,
the study based on detailed simulation\cite{ZHHsim} predicts
$\sim 10\mathrm{\%}$ precision will be expected.
This requires the estimation of the radiative correction.
The genuine weak correction is a few \% for the probable
values of $W, M_H$\cite{ZHH}. 

In this paper, preliminary results for the
two processes, $e^+e^-\rightarrow e^+e^-H$ 
and $e^+e^-\rightarrow \nu\bar{\nu}\gamma$
are presented based on the works by the authors.

\section{System}

The calculation is done using the automated system for
the perturbative calculation, GRACE-LOOP.
The detailed description is found in \cite{GRACEL} and
only a few features are given here.

\subsection{5-point functions}

In order to calculate the radiative correction for
two-to-three processes, one needs 5-point functions.
Since a set of 5 vectors are linearly dependent in 4-dimensional
space, an identity for the Gram determinant leads
to the reduction of a 5-point function into a sum
of lower point functions\cite{fiveA}.
The explicit implementation of the reduction formula
is not unique. For example, see \cite{fiveB} for the 
similar implementation by which the one-loop QED
correction is calculated for $ e^+e^-\rightarrow \mu^+\mu^-H$.
We have examined several methods and found that
the following is appropriate in order to make
the produced symbolic formulae {\it relatively} short.

\[
I_5=\int\frac{N}{D_0D_1D_2D_3D_4}
\]
\begin{equation}
\quad=\sum G_{\mu\nu\cdots\sigma}\int
\frac{\ell^{\mu} \ell^{\nu} \cdots \ell^{\sigma} }{D_0D_1D_2D_3D_4}
\label{eq:five}\end{equation}
where
the numerator is a rank $M$ tensor of $\ell$,
$D_0=\ell^2+X_0$ and $D_j=\ell^2+2 r_j\ell +X_j\ (j=1,2,3,4)$.
We can define the 'metric tensor' by
\begin{equation}
g^{\mu\nu}=\sum_{ij} r_i^{\mu}(A^{-1})_{ij}r_j^{\nu},\qquad A_{ij}=r_ir_j
\label{eq:fivea}\end{equation}
to obtain the identity
\[
\ell^{\mu}=g^{\mu\nu}\ell_{\nu}
\]
\begin{equation}
\quad=\sum_{ij}\frac{1}{2} r_i^{\mu}(A^{-1})_{ij}(D_j-D_0+X_0-X_j).
\label{eq:fiveb}\end{equation}
Substituting the above identity into Eq.(\ref{eq:five}), the
numerator becomes
\begin{equation}
N=\sum_{\alpha=0}^4 E_\alpha(\ell)D_\alpha + F.
\label{eq:fivec}\end{equation}
The first term of Eq.(\ref{eq:fivec}) turns to be
box integrals whose numerator is a rank $M-1$ tensor of $\ell$.
The second term of Eq.(\ref{eq:fivec}) becomes a scalar 5-point
function.
This can be reduced by the identity
\begin{equation}
1 = \sum_{\alpha=0}^4 (a_\alpha+b_\alpha r_j\ell)D_\alpha
\label{eq:fived}\end{equation}
which can be obtained from $D_0-X_0=\ell_{\mu}g^{\mu\nu}\ell_{\nu}$.

\begin{table}[hbt]
\caption{The number of Feynman diagrams}  
\label{table:1}
\renewcommand{\tabcolsep}{0.8pc} % enlarge column spacing
\renewcommand{\arraystretch}{1.2} % enlarge line spacing
\begin{tabular}{|c|r|r|}
\hline
process & tree & 1-loop \\ \hline
$e^+e^-\rightarrow \nu\bar{\nu}H$ & 12(2) & 1350(249) \\ \hline
$e^+e^-\rightarrow t\bar{t}H$ & 12(6) & 2327(758) \\ \hline
$e^+e^-\rightarrow ZHH$ & 27(6) & 5417(1597) \\ \hline
$e^+e^-\rightarrow e^+e^-H$ & 42(2) & 4470(510) \\ \hline
$e^+e^-\rightarrow \nu\bar{\nu}\gamma$ & 10(5) & 1099(331) \\ \hline
\end{tabular}\\[2pt]
\hfill full set(production set) 
\end{table}

\subsection{Non-linear gauge}

We have implemented the non-linear gauge fixing 
defined by the following Lagrangian\cite{nlg,GRACEL}.

\mathindent 0pt

\begin{equation}
L_{GF}=
 - \frac{1}{\xi_w} F^+ F^-  
 -\frac{1}{2\xi_z}(F^Z)^2  
 -\frac{1}{2\xi}(F^A)^2  
\end{equation}

\[
F^{\pm}=
\left( \partial^{\mu} \mp i e \tilde{\alpha} A^{\mu}
       \mp i \frac{ec_W}{s_W} \tilde{\beta} Z^{\mu}
\right) W_{\mu}^{\pm} 
\]
\begin{equation}
+ \xi_w \left( M_W \chi^{\pm} +\frac{e}{2 s_W} \tilde{\delta} H \chi^{\pm}
\pm i \frac{e}{2 s_W}\tilde{\kappa} \chi_3 \chi^{\pm}
\right)
\end{equation}
\begin{equation}
F^Z=\partial^{\mu}Z_{\mu} 
+ \xi_z\left( M_Z \chi_3  +\frac{e}{2s_Wc_W}\tilde{\epsilon} H \chi_3 \right) 
\end{equation}
\begin{equation}
F^A=\partial^{\mu}A_{\mu} 
\end{equation}

The main reason we use this gauge fixing is to check the
output of the system.  In the calculation based on an
automated system, the diagnostic stage is highly important.
Among many check items, e.g., renormalizability, infrared
stability, etc., the gauge invariance is most powerful
check.

In order to keep the loop integral simple and stable,
the numerator of vector propagators is to be $g^{\mu\nu}$.
Under this restriction, the gauge check is not possible
in the conventional $R_{\xi}$ gauge.
Sometimes, the non-linear gauge fixing was used
to reduce the number of diagrams: For instance, the choice of
$\tilde{\alpha}=1$ removes $\gamma W\chi$ vertex. \hfill
We take $\xi_w=\xi_z=\xi=1$ while $\tilde{\alpha}, \tilde{\beta},
\tilde{\delta}, \tilde{\epsilon}, \tilde{\kappa}$ are
arbitrary.  Then the number of diagrams is larger than 
that in the linear gauge fixing.

GRACE system generates all possible Feynman diagrams for a
specified process. We call them 'full set'.
Using the full set and in quadruple precision
we repeat the computation of the cross section  at
a few phase space points for a different set of numerical
values of gauge parameters. 
When they agree within reasonable digits,
we can confirm the performance of the system.
Then we discard the diagrams containing scalar-electron couplings
to define 'production set'.
The cross section of the production set is integrated
in the phase space in double precision (sometimes in quadruple precision
for the check).
The numbers of diagrams are shown in Table 1. 
For the 1-loop, the counter terms are counted also.

\section{Results}

\subsection{ $ e^+e^-\rightarrow e^+e^-H$ }

\begin{figure}[htb]
\includegraphics{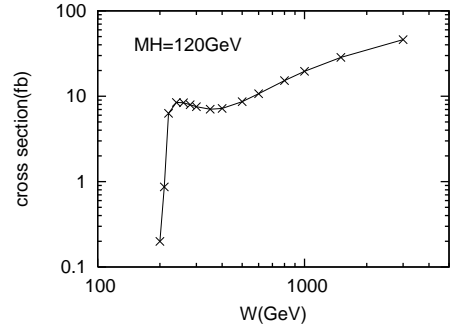}
\caption{Tree cross section for $ e^+e^-\rightarrow e^+e^-H$.}
\label{fig:fig1}
\end{figure}

This channel is associated with the dominant Higgs production processes
$e^+e^-\rightarrow \nu_e\bar{\nu}_eH$.
Higgsstrahlung $e^+e^-\rightarrow ZH$ and
the $Z$-fusion process contribute to this process.
As is shown in Fig.\ref{fig:fig1}, the tree cross section
rises quickly after the $ZH$ threshold and has
the values of $O(10)\mathrm{fb}$.

The parameter set for the calculation is as follows:
$M_W=80.3766\mathrm{GeV}$, \\
$M_Z=91.1876\mathrm{GeV}$, 
$\Gamma_Z=2.4956\mathrm{GeV}$ 
$M_H=120\mathrm{GeV}$, 
$m_t=174\mathrm{GeV}$, 
$W= 200\sim3000\mathrm{GeV}$,
$k_{cut}=0.05E$
The width of $Z$ only appears at resonant poles.

In Fig.2, the EW correction is shown. 
Here, $\delta(W)$ is the fraction of 
the deviation from the tree cross section
after subtracting the well-known QED correction.
For the parameter set used here, $\Delta r=2.55\mathrm{\%}$.
In the so-called $G_{\mu}$-scheme, 
$\delta(W)^G=\delta(W)-3\Delta r$.
$G_{\mu}$-scheme explains the major part of EW correction in
lower energy region, but in higher energy region.

\begin{figure}[htb]
\includegraphics{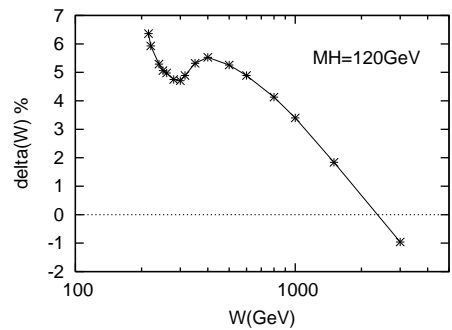}
\caption{Weak correction $\delta(W)$ for $ e^+e^-\rightarrow e^+e^-H$.}
\label{fig:fig2}
\end{figure}

\subsection{ $ e^+e^-\rightarrow \nu\bar{\nu}\gamma$ }

The single photon signal contributes to many SUSY
particle study.
The precise understanding of the radiative correction
in the standard model is essential for such a study.
Also this channel serves for the neutrino counting.
The physical parameters are the same as for $ e^+e^-\rightarrow e^+e^-H$.
For the radiated photons, we have applied the so-called
OPAL cut, i.e., 
$p_T(\gamma)>0.05E, 15^{\circ}<\theta(\gamma)<165^{\circ}$.

\begin{figure}[htb]
\includegraphics{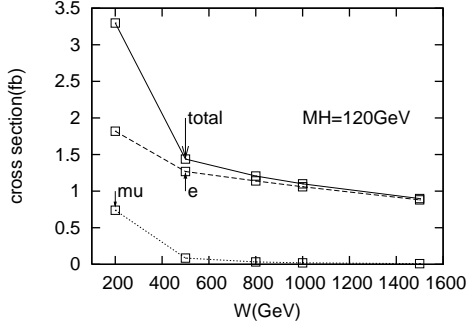}
\caption{Tree cross section for $ e^+e^-\rightarrow \nu\bar{\nu}\gamma$.}
\label{fig:fig3}
\end{figure}

As is shown in Fig.\ref{fig:fig3}, the tree cross section
is about $1\mathrm{fb}$ and the $\nu_e$-contribution
dominates for $W>500\mathrm{GeV}$.

\begin{figure}[htb]
\includegraphics{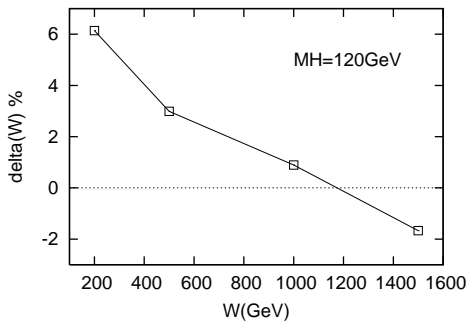}
\caption{Weak correction $\delta(W)$ 
for  $e^+e^-\rightarrow \nu\bar{\nu}\gamma$.}
\label{fig:fig4}
\end{figure}

The energy dependence of EW correction differs
between $\nu_e$-contribution and that of $\nu_{\mu}$.
For instance, at $W=1.5\mathrm{TeV}$ the correction for $\nu_{\mu}$
is 4 times larger than that of $\nu_e$.
However, as $\nu_{\mu}$ term is small in high energy, the
weak correction is determined mostly by the $\nu_e$-contribution.
The correction is shown in Fig.\ref{fig:fig4}.
The definition of the $\delta(W)$ is the same as in the last
subsection. The energy dependence shows common structure to
other processes.
The photon energy and angle distributions 
will be presented in the coming publication.

\section{Final remarks}

The study of higher order effects is important for the
future study of Higgs and SUSY particles. It has
been demonstrated that the GRACE-LOOP system serves 
well for such a study.
We have calculated the full EW radiative corrections
for two processes, $ e^+e^-\rightarrow e^+e^-H$
and $e^+e^-\rightarrow\nu\bar{\nu}\gamma$. 
At $W=500\mathrm{GeV}$ and for $M_H=120\mathrm{GeV}$, 
the genuine EW correction is 5\% for the former 
and 3\% for the latter.
The presented results are preliminary, and the
detailed study will appear in the forthcoming
publications.

%-----------------------------------------------------
\vspace{6mm}
\noindent {\bf Acknowledgment} \hfill
We gratefully acknowledge the participation of
Genevi\`{e}ve B\'{e}langer throughout this project and 
would like to thank her help and comments.
We would like to thank D.Perret-Gallix who created the
series of workshop in 1990 and still keeps continuous
interest and encouragement on our works.
This work was supported in part by the Japan Society 
for Promotion of Science under Grant-in-Aid for
Scientific Research B(No.14340081), PICS 397 of
the French National Center for Scientific Research(CNRS).

%-----------------------------------------------------

\end{document}